\documentclass[aps,prd,twocolumn,amsmath,amssymb,floatfix]{revtex4-2}
\usepackage{graphicx}
\usepackage{rotating}
\usepackage{color}
\usepackage{soul}
\usepackage{hyperref}
\usepackage{threeparttable}
\usepackage{natbib}
\usepackage{longtable}
\usepackage{appendix}
\usepackage{array}
\usepackage[caption=false]{subfig}
\usepackage{tabularx}
\usepackage[utf8]{inputenc}

\newcommand{\blue}[1]{\textcolor{black}{#1}}
\newcommand{\cyan}[1]{\textcolor{black}{#1}}

\begin{document}

\title{Constraints from big bang nucleosynthesis and cosmological observations on varying Higgs vacuum expectation value}

\author{Hongrui Feng$^{1}$}

\author{Yudong Luo$^{2,3}$}
\email[Contact author: ]{ydongluo@ibs.re.kr}

\author{Toshitaka Kajino$^{1, 4, 5}$}
\email[Contact author: ]{kajino@buaa.edu.cn}

\author{Bao-Hua Sun$^{1}$}
\email[Contact author: ]{bhsun@buaa.edu.cn}

\author{Tatsushi Shima$^{6}$}

\affiliation{$^1$School of Physics, Peng Huanwu Collaborative Center for Research and Education, and International Research Center for Big-Bang Cosmology and Element Genesis, Beihang University, Beijing 100191, China}
\affiliation{$^2$Center for Exotic Nuclear Studies, Institute for Basic Science, Daejeon 34126, Korea}
\affiliation{$^3$School of Physics and Kavli Institute for Astronomy and Astrophysics, Peking University, Beijing 100871, China}
\affiliation{$^4$Graduate School of Science, The University of Tokyo, Tokyo 113-0033, Japan}
\affiliation{$^5$National Astronomical Observatory of Japan, Tokyo 181-8588, Japan}
\affiliation{$^6$Research Center for Nuclear Physics, The University of Osaka, Osaka 567-0047, Japan}

\date[]{}

\begin{abstract}
The Higgs vacuum expectation value (VEV) alters both the electroweak and strong interaction rates.
We study both effects on big bang nucleosynthesis (BBN) and seek concordance between observed primordial abundances of light elements and cosmological constraints from CMB fluctuations and anisotropies.
We find a strong negative correlation between primordial $^4$He abundance and the Higgs VEV. 
Consequently, using a 1.58\% uplift of the Higgs VEV during BBN from the current value, an agreement can be achieved among the new primordial $^4$He abundance, observed by the EMPRESS group, Deuterium abundance determined from absorption lines in the Lyman-alpha forest along the line-of-sight of high-redshift quasars\blue{, and the lithium abundance on the Spite plateau.  This agreement requires a baryon-to-photon ratio different from the CMB determination, indicating the need for nonstandard cosmological evolution between BBN and recombination epochs.} 
\blue{We also demonstrate that a 0.2\% uplift in the Higgs VEV can partially alleviate the ``cosmic Li problem,'' keeping consistency with the CMB with remaining discrepancy potentially accounted for by stellar depletion mechanisms.} 
\end{abstract}

\maketitle

\begin{center}
\small
Journal reference: Phys. Rev. D 114, 083017 (2026)\\
DOI: \href{https://doi.org/10.1103/6zff-cqtk}{10.1103/6zff-cqtk}
\end{center}

\section{Introduction}
Big bang nucleosynthesis (BBN) describes the first 0.01--100\,s of our Universe, during which light elements, such as $^4$He, $^2$H (hereafter, D), $^3$He, and a trace amount of $^7$Li were created. Standard BBN (SBBN) is based on the standard $\Lambda$ cold dark matter ($\Lambda$CDM) model under the assumption of a homogeneous and isotropic Universe~\cite{wagoner1973big, Olive:1989xf, Olive:1999ij, Fields:2006bzp, Coc:2014oia, bertulani2016frontiers, cyburt2016big, fields2020big, Cyburt:2001pq, cyburt2003primordial}. 
In addition to the primordial elemental abundances, the $\Lambda$CDM model effectively accounts for other cosmological and astrophysical observations, including the CMB fluctuations and anisotropies, baryon acoustic oscillations (BAO), and the large-scale structure of galaxies~\cite{Peebles:1970ag, Voit:2004ah, Costa:2021jsk, SDSS:2023tbz, DESI:2024mwx}. \par
However, the standard $\Lambda$CDM model has several issues. The baryon density in BBN is described by a single model parameter, $\Omega_b h^2$, which can be independently determined from the CMB~\cite{WMAP:2012nax, ade2016planck, aghanim2020planck}. Using this parameter, the BBN predictions of D and $^{4}$He are in excellent agreement with observations~\cite{Pettini:2012ph, Cooke:2013cba, cooke2018one} and~\cite{Izotov:2007ed, Cooke:2018qzw, hsyu2020phlek, aver2015effects}, respectively. In contrast, the situation for lithium remains problematic.
Spite and Spite~\cite{1982A&A...115..357S, Spite1982Nature} found that many metal-poor halo stars share a nearly constant lithium abundance, and proposed that this ``Spite plateau'' reflects the primordial lithium abundance (see~\cite{sbordone2010metal} for recent observation). However, adopting the Planck value $\Omega_b h^2 = 0.02237\pm0.00015$~\cite{aghanim2020planck}, the ``Spite plateau'' is lower than the standard BBN prediction by about a factor of 3~\cite{fields2011the}. \cyan{A homogeneous nonlocal thermodynamic equilibrium analysis of metal-poor stars selected from the LAMOST survey suggests that the Spite plateau extends to lower metallicities with a mild positive slope, while remaining significantly below the SBBN prediction~\cite{Yan:2026vls}.} Studies like Korn {\it et al.}~\cite{korn2006probable} and Fu {\it et al.}~\cite{2015MNRAS.452.3256F} argue that lithium in metal-poor stars has been altered by stellar evolutionary processes, including diffusion, turbulent mixing, and pre-main-sequence lithium depletion. 
\blue{More recently, Fields and Olive~\cite{Fields:2022mpw} argued that the lithium discrepancy may be fully explained by stellar depletion, based on stringent constraints on the $^{6}\mathrm{Li}/^{7}\mathrm{Li}$ ratio in metal-poor halo stars.} 
Besides this ``cosmic Li problem,'' there is significant tension in the measurement of Hubble constant $H_0$, with a 4--6$\sigma$ discrepancy existing between late-time local Hubble flow and early-time indirectly inferred value from CMB~\cite{Riess:2019cxk,aghanim2020planck,Verde:2019ivm}. \par
\blue{The possibility of cosmological evolution of fundamental constants can be traced back to Dirac's large number hypothesis~\cite{Dirac:1937ti}. Dixit and Sher~\cite{Dixit:1987at} first investigated the impact of a varying Higgs vacuum expectation value (VEV) on big bang nucleosynthesis by considering the variation of the Fermi constant. Subsequent studies have explored the effects of Higgs VEV variation on weak interactions and the neutron-proton conversion processes during the BBN epoch \cite{burns2024constraints,meyer2024improved,PhysRevD.47.4774,ichikawa2002constraining,yoo2003big,muller2004nucleosynthesis}. Meanwhile, the impact of quark mass variations on BBN through modifications of nuclear binding energies and reaction rates has been investigated \cite{cheoun2011time,mori2019roles}.}
Dolan {\it et al.}~\cite{Dolan:1973qd}, Weinberg~\cite{Weinberg:1974hy}, and Graham {\it et al.}~\cite{Graham:2015cka} described the dynamical onset of a nonzero Higgs VEV during inflation. 
These considerations therefore motivate the hypothesis that the Higgs VEV in the early Universe may have differed from its present-day value. 
Fung {\it et al.}~\cite{fung2021axi} proposed such an ``axi-Higgs'' cosmology model, in which an ultralight axion couples to the Higgs field and slightly increases the Higgs VEV in the early universe. Since the axion mass are light, the decouple time is late. As the axion starts rolling near or after recombination, the Higgs VEV decreases toward its present value and subsequently oscillates with a highly suppressed amplitude at late times.  Recently, Burns {\it et al.}~\cite{burns2024constraints} and Meyer {\it et al.}~\cite{meyer2024improved} discussed the constraint on the Higgs VEV in the BBN era, imposed by the primordial $^4$He determination from the EMPRESS group~\cite{matsumoto2022empress}.\par
\blue{Although these studies have provided important insights into the effects of a varying Higgs VEV during BBN, some relied on linear extrapolations from sensitivity studies rather than self-consistent BBN network calculations. Moreover, none simultaneously incorporated the correlated electroweak and strong-interaction effects of Higgs VEV variation into a self-consistent network while carefully accounting for shifts in nuclear resonance energies.}
\blue{In this work, we perform a self-consistent BBN analysis with a varying Higgs VEV by incorporating both the weak-interaction effects and the nuclear effects induced by quark mass variations. We solve the BBN evolution using a nuclear reaction network, including the possible impact of shifts in nuclear resonance energies. By comparing the predicted primordial abundances with the latest observational determinations of $^4$He, D, and Li, we derive constraints on the Higgs VEV variation during the BBN epoch and investigate its implications for the cosmic Li problem and the baryon density.}\par
\blue{Since the electron mass is proportional to the Higgs VEV in the Standard Model~\cite{luu2023axion}, a varying Higgs VEV provides a natural framework to investigate its implications for the Hubble tension.} Hart {\it et al.}~\cite{hart2020updated,Hart:2017ndk} and Seto {\it et al.}~\cite{Seto:2024cgo} suggested that the Hubble tension could be alleviated by allowing a slight increase in the electron mass $m_e$ during recombination.  
By contrast, since the fine-structure constant is independent of the Higgs VEV, the VEV variation considered here does not conflict with the stringent observational bounds on the time variation of the fine-structure constant derived from absorption spectra of high-redshift quasars~\cite{Songaila:2014fza, Wei:2024trz}. \par
This article is arranged as follows. In Sec.~\ref{models}, we examine how variations in the Higgs VEV during the BBN epoch alter fundamental constants, namely quark masses, the electron mass, and the Fermi constant. We further explore the resulting impacts on both electroweak and strong interactions, as well as resonant thermonuclear reactions. In Sec.~\ref{results}, we analyze the dependence of primordial $^4$He, D, and $^7$Li abundances to the Higgs VEV. We report constraints on the Higgs VEV and the baryon density in the early-Universe from primordial $^4$He and D observations. By incorporating newly observed $^4$He abundance~\cite{matsumoto2022empress}, we could find agreement among primordial $^4$He, D, and $^7$Li abundances. Furthermore, when considering the Lithium depletion mechanism in metal-poor stars~\cite{korn2006probable}, we demonstrate that a small uplift of Higgs VEV during the BBN epoch would provide a possible cosmological solution to the ``cosmic Li problem.'' In Sec.~\ref{conclusion}, we summarize our findings.\par

\section{Theoretical methods}
\label{models}
Fundamental particles acquire their mass through their coupling to the Higgs field. If the Higgs VEV at the BBN epoch, $v_\mathrm{BBN}$, differs from its present-day value,$v_0$, it scales the boson, electron, and quark masses. This alters the weak interaction rates $\lambda(n \rightarrow p)$ and $\lambda(p \rightarrow n)$, which are crucial for determining the $n/p$ ratio during weak interaction decoupling~\cite{burns2024constraints, meyer2024improved}.
The Fermi constant $G_F$, which describes the effective coupling strength of the weak interaction at low energies, originates from integrating out the massive weak boson. It is inversely proportional to the Higgs VEV squared. Thus, we consider that, during BBN, $G_F = 1/\sqrt{2} v_{\rm BBN}^2$~\cite{PhysRevD.47.4774}. 
The mass of charged leptons and quarks is generated from spontaneous symmetry breaking of the electroweak theory via Yukawa interactions. So electron mass $m_e \propto v_{\rm BBN}$, and the up-down quark mass difference scales as $\delta m_q \propto v_{\rm BBN}$. This directly alters the neutron-proton mass difference $Q_{np}$~\cite{fung2021axi,muller2004nucleosynthesis,ichikawa2002constraining,yoo2003big},
\begin{equation}
    \label{Qnp_dependence}
    Q_{np} = \left[-0.76+2.05\frac{v_\mathrm{BBN}}{v_0}\right]\,\text{MeV}.
\end{equation}

\begin{figure*}[t]
    \centering
    \includegraphics[width=1.0\textwidth]{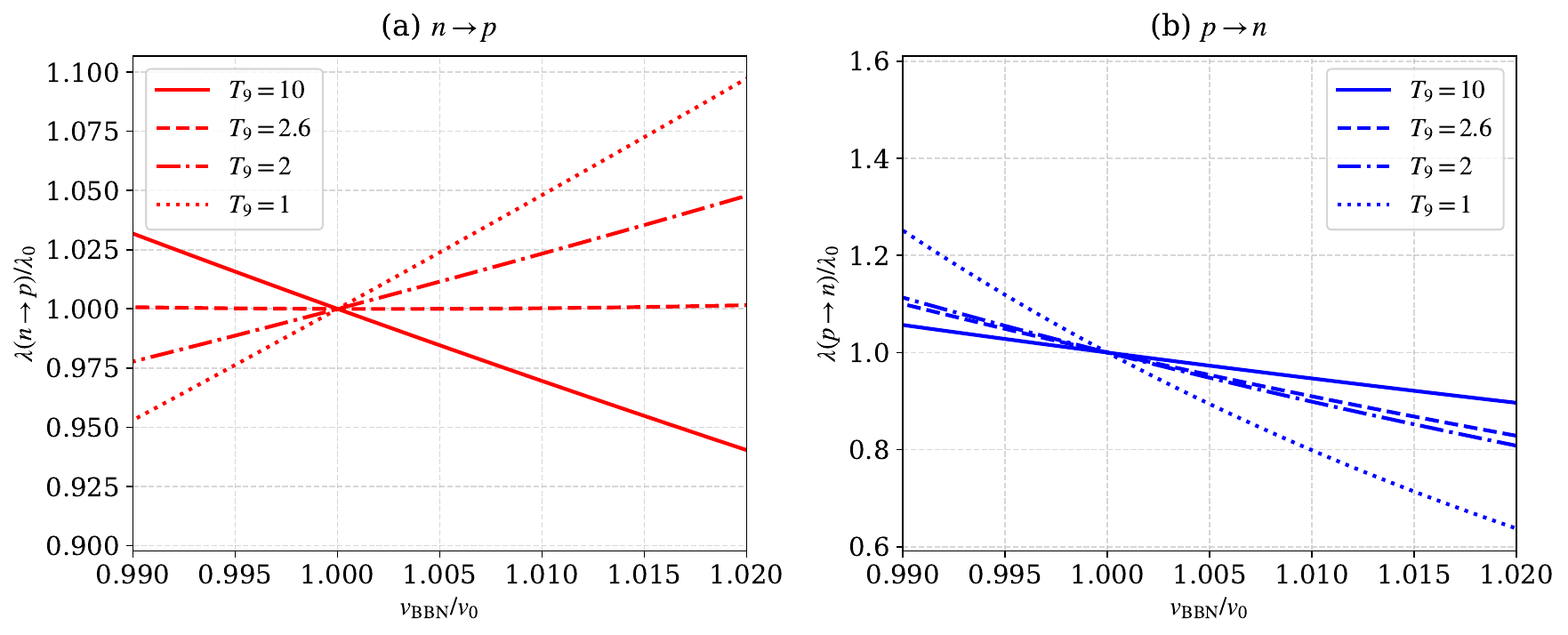}
    \caption{Higgs VEV's impact on weak interaction: relative change of (a) $n \rightarrow p$ and (b) $p \rightarrow n$ rates. Assuming no change in the Higgs VEV, the weak transition rates ($\lambda_0(n \rightarrow p), \lambda_0(p \rightarrow n)$) are: $(0.81\,\mathrm{s^{-1}}, 0.18\,\mathrm{s^{-1}})$ at \blue{$T_9 = 10$}; $(3.0\times10^{-3}\,\mathrm{s^{-1}}, 8.3\times10^{-6}\,\mathrm{s^{-1}})$ at \blue{$T_9 = 2.6$}; $(1.7\times10^{-3}\,\mathrm{s^{-1}}, 7.4\times10^{-7}\,\mathrm{s^{-1}})$ at \blue{$T_9 = 2$};  and $(1.1\times10^{-3}\,\mathrm{s^{-1}}, 8.0\times10^{-11}\,\mathrm{s^{-1}})$ at \blue{$T_9 = 1$}.}
    \label{fig_higgs_wrate}
\end{figure*}

In Fig.~\ref{fig_higgs_wrate}, we show the impact of Higgs VEV on $n \leftrightarrow p$ conversion rates. 
Following weak interaction freeze-out at \blue{a temperature $T_9\equiv \frac{T}{10^9\,\mathrm{K}}=10$}, both the $n \rightarrow p$ and $p \rightarrow n$ transition rates decline as $v_\mathrm{BBN}$ increases. The $n/p$ ratio is thus primarily governed by the mass difference $Q_{np}$, which scales positively with $v_\mathrm{BBN}$. Consequently, an increased Higgs VEV ultimately suppresses the number of free neutrons.
At temperatures below \blue{$T_9 = 2.6$}, the effect of Higgs VEV on $\lambda(n \rightarrow p)$ becomes negative, and the $\lambda(p \rightarrow n)$ rate falls rapidly out of relevance.
At \blue{$T_9 = 2$}, D and $^3$He decouple, and at \blue{$T_9 = 1$}, the Deuterium bottleneck is breached. At this stage, the $n \rightarrow p$ rate is already very close to the free neutron $\beta$-decay rate corresponding to the measured neutron lifetime~\cite{navas2024review}. 
Here, the dominant effect of $v_{\mathrm{BBN}}$ on the $n \rightarrow p$ transition arises from a larger $Q_{np}$, which now accelerates neutron decay. Additional modifications come from changes in the electron mass $m_e$ and the Fermi constant $G_F$. Specifically, a larger $m_e$ reduces the accessible phase space for electrons, and the quadratic dependence of the interaction rate on $G_F$ suppresses the decay rate. Both factors contribute to a smaller neutron destruction rate for larger $v_{\mathrm{BBN}}$. The overall correlation of neutron survival time on Higgs VEV is negative; thus, a larger $v_{\mathrm{BBN}}$ would lead to a lower number of free neutrons, lowering the primordial $^4\mathrm{He}$ abundance.
\par
As for the thermonuclear reactions induced by electromagnetic and strong forces, the Higgs VEV dependence arises from the averaged quark mass $m_q = \left(m_u+m_d\right) / 2 \propto v_\mathrm{BBN}$~\cite{fung2021axi}. Variations in the Higgs VEV affect the reaction dynamics in three ways: the change in exit-channel kinetic energy affects reaction cross sections; the resulting shift in reaction $Q$-values alters the balance between forward and reverse reactions; and the modification of resonance energy levels substantially impacts reaction rates within the BBN energy range~\cite{cheoun2011time, mori2019roles}. \par
The dependence of nuclear binding energies $B$ on the averaged quark mass, and by extension to $v_{\mathrm{BBN}}$, is described by a dimensionless coefficient $K$ taken from~\cite{flambaum2007dependence}, yielding $\delta B/B = K (\delta m_q/m_{q0}) = K(v_{\mathrm{BBN}}/v_0 -1)$. 
When the Higgs VEV is altered, the resulting change in the reaction $Q$-value is given by,
\begin{equation}
    \delta Q = \sum_i B_i K_i \left(\frac{v_{\mathrm{BBN}}}{v_0}-1\right) - \sum_j B_j K_j \left(\frac{v_{\mathrm{BBN}}}{v_0}-1\right), 
\end{equation}
with $i$ and $j$ representing the entrance-channel and exit-channel nuclei, respectively. The cross-section dependence arises because the reaction $Q$-value contributes to the exit-channel kinetic energy. This alters both the final-state velocity and the exit-channel Coulomb-barrier penetrability via the Gamow term. At BBN temperatures, the changes of cross section $\delta \sigma$ for charged-particle emission and radiative capture reactions scale linearly with the change of Higgs VEV, 
\begin{equation}
    \frac{\delta \sigma}{\sigma_0} = k_{\sigma v} \left(\frac{v_{\mathrm{BBN}}}{v_0}-1\right),
\end{equation}
where the proportionality coefficients $k_{\sigma v}$ are tabulated in Table~\ref{table_rate_ratio}~\cite{cheoun2011time,mori2019roles}. Note that for the $\mathrm{^1H}(n,\gamma)\mathrm{^2H}$ reaction, the cross-section is further influenced by the virtual energy level at 0.07\,MeV~\cite{dmitriev2004cosmological}. Furthermore, since neutron emission reactions lack an exit-channel Coulomb barrier, the cross sections for the $^2\mathrm{H}(d,n)^3\mathrm{He}$ and $^3\mathrm{H}(d,n)^4\mathrm{He}$ reactions are insensitive to $Q$-value variations, yielding $k_{\sigma v}=0$.
\begin{table*}[t]
    \caption{Proportionality coefficients $k_{\sigma v}$ relating the Higgs VEV to the changes of cross section for charged-particle emission and radioactive capture reactions. 
    For neutron-emission reactions, $k_{\sigma v}=0$.
    }
\label{table_rate_ratio}
\begin{ruledtabular}
\begin{tabular}{cccc}
\multicolumn{2}{c}{Charged-particle emission} & \multicolumn{2}{c}{Radiative capture} \\
\hline
Reaction & $k_{\sigma v}$ & Reaction & $k_{\sigma v}$ \\
\hline
$\mathrm{^3He}(n,p)\mathrm{^3H}$ & $-0.326$ & $\mathrm{^1H}(n,\gamma)\mathrm{^2H}$ & $-11.3$ \\
$\mathrm{^7Li}(p,\alpha)\mathrm{^4He}$ & $-0.152$ & $\mathrm{^2H}(p,\gamma)\mathrm{^3He}$ & $-4.84$ \\
$\mathrm{^2H}(d,p)\mathrm{^3H}$ & $-1.07$ & $\mathrm{^3H}(\alpha,\gamma)\mathrm{^7Li}$ & $-19.6$ \\
$\mathrm{^3He}(d,p)\mathrm{^4He}$ & $-0.597$ & $\mathrm{^3He}(\alpha,\gamma)\mathrm{^7Be}$ & $-31.2$ \\
\end{tabular}
\end{ruledtabular}
\end{table*}
Since the change of cross section induced by kinetic energy in the exit channel is temperature independent, the resulting modification in the reaction rate is also temperature independent. Notably, all of the reaction rates decrease as $v_{\mathrm{BBN}}$ increases. This arises from the fact that heavier nuclei are generally more sensitive to the changes of quark mass.\par

\begin{figure*}[t]
    \centering
    \includegraphics[width=1.0\textwidth]{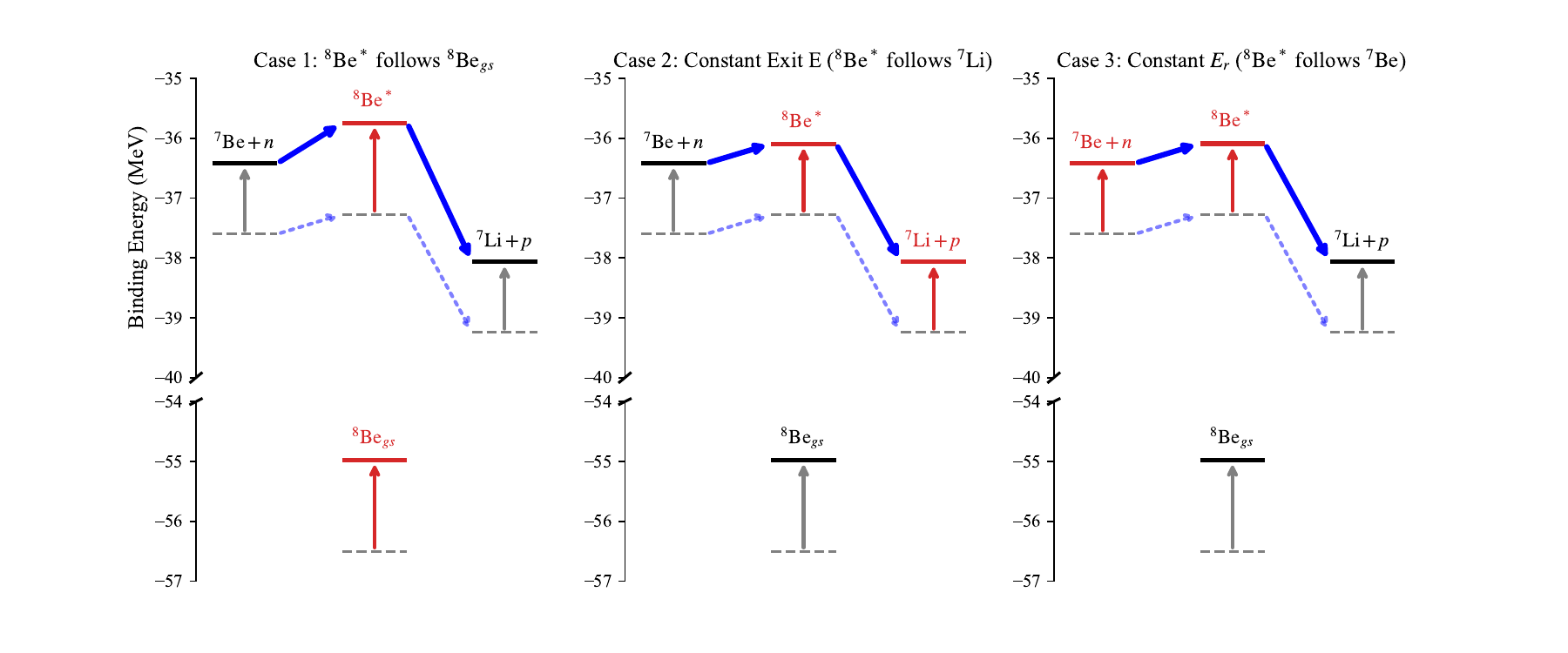}
    \caption{Illustration of the three difference responses to a change in $m_q$ of the resonance energy for the reaction $\mathrm{^7Be}(n,p)\mathrm{^7Li}$. Case 1: Variations in binding energy of excited states is the same as the ground state. Case 2: The resonance height does not change in the reverse reaction. Case 3: No shift in the resonance energy. The dashed lines represent the energy levels before the change, and the solid lines represent the energy levels after the change.}
    \label{resonance_illustration}
\end{figure*}
In BBN, three resonant reactions $\mathrm{^3H}(d,n)\mathrm{^4He}$, $\mathrm{^3He}(d,p)\mathrm{^4He}$ and $\mathrm{^7Be}(n,p)\mathrm{^7Li}$ significantly affect the primordial D and $^7$Li abundances. The first two reactions are narrow resonance reactions~\cite{caughlan1988thermonuclear, descouvemont2004compilation}, while the third reaction exhibits a broad resonance structure~\cite{hayakawa2021constraining,damone20187}. Their cross sections as a function of the center-of-mass energy $E$ can be described through the Breit-Wigner formula~\cite{mori2019roles,yao2025implication}, where the resonance energy $E_r$ depends on the Higgs VEV. 
Since $E_r$ is determined by the position of the excited state of the compound nucleus relative to the entrance-channel threshold, variations in the quark mass $m_q$ can modify the reaction rate through shifts in both the nuclear binding energies and the resonance position. 
However, the response of compound-nucleus excited states to a change in $m_q$ remains uncertain. Here we explore three scenarios, similarly to Mori et al.~\cite{mori2019roles}: in case 1, the resonance-state energy shifts in the same way as the ground-state binding energy; in case 2, the resonance shift is constrained by requiring the resonance strength in the reverse reaction to remain unchanged; and in case 3, the resonance energy $E_r$ is assumed to remain fixed, corresponding to a resonance-state shift that follows the entrance-channel threshold. 
Figure~\ref{resonance_illustration} illustrates these three cases for the $\mathrm{^7Be}(n,p)\mathrm{^7Li}$ reaction, where the red arrows indicate the corresponding shifts in the energy levels.\par
For narrow resonance reactions $\mathrm{^3H}(d,n)\mathrm{^4He}$ and $\mathrm{^3He}(d,p)\mathrm{^4He}$, the changes in $E_r$ relative to the unperturbed resonance energy $E_{r0}$ directly alter the reaction rates~\cite{cheoun2011time,PhysRevD.70.023505},
\begin{equation}
    \label{narrow_res}
    \begin{aligned}
    \frac{N_A \langle \sigma v \rangle}{\left[N_A \langle \sigma v \rangle\right]_0}
    &= \left[1 + k_{\sigma v}\left(\frac{v_{\mathrm{BBN}}}{v_0} -1\right)\right] \\
    &\quad \times \left[\frac{1+\left(E_0-E_r\right)^2/\left(\Gamma_r/2\right)^2}{1+\left(E_0-E_{r0}\right)^2/\left(\Gamma_r/2\right)^2}\right].
    \end{aligned}
\end{equation}
Here $E_0 = E_G^{\frac{1}{3}}\left(kT/2\right)^{\frac{2}{3}}$ is the energy peak of the Gamow window with $E_G = \frac{2 \pi^2 e^4}{\hbar^2} Z_1^2 Z_2^2 \mu$, where $Z_1$ and $Z_2$ represent the charges of the incoming nuclei, and $\mu$ is the reduced mass.
\begin{figure*}[t]
    \centering
    \includegraphics[width=1.0\textwidth]{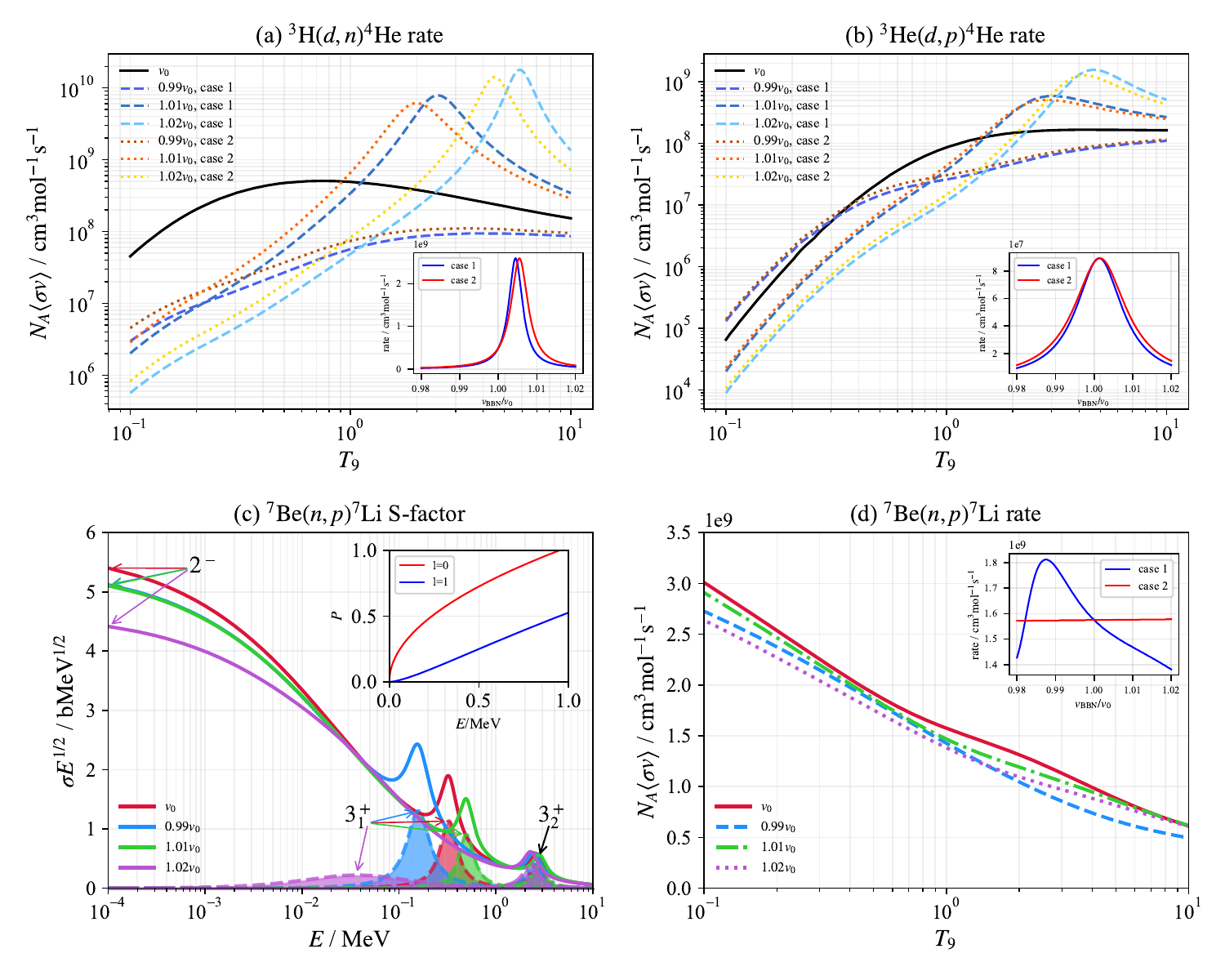}
    \caption{Effects of variations in the Higgs VEV on resonance reaction properties. Panels (a) and (b) correspond to reaction rates of narrow resonance reactions $\mathrm{^3H}(d,n)\mathrm{^4He}$ and $\mathrm{^3He}(d,p)\mathrm{^4He}$, with respect to $v_{\mathrm{BBN}}$. The solid curves represent the original reaction rates, the dashed and dotted curves correspond to cases 1 and 2 of resonance energy level variation. The lower-right insets show the resulting change in the reaction rates at the BBN-sensitive temperature \blue{$T_9 = 1$}. Panels (c) and (d) show, respectively, the S-factor and reaction rate for the broad-resonance reaction $\mathrm{^7Be}(n,p)\mathrm{^7Li}$ in case 1. In panel (c), the zero-energy S-factor corresponds to the contribution from the $2^-$ resonant state, while shaded regions highlight the shift of the two $3^+$ resonance components with $v_{\mathrm{BBN}}$. The solid curve represents the total S-factor. The upper-right inset shows the energy dependence of the penetration factor $P(E;l)$. Panel (d) presents the corresponding variation in the reaction rate; its inset shows the change in the reaction rate at \blue{$T_9 = 1$} for resonance cases 1 and 2.}
    \label{fig_resonance}
\end{figure*}
As shown in panels (a) and (b) of Fig.~\ref{fig_resonance}, $E_0$ exhibits a positive correlation with $v_\mathrm{BBN}$. The reaction rate is enhanced when the resonance energy $E_r$ lies close to $E_0$, and it decreases sharply once $E_r$ moves outside the BBN-relevant Gamow window.\par
The cross section of the broad resonance reaction $\mathrm{^7Be}(n,p)\mathrm{^7Li}$ is characterized by the contributions of multiple resonant states in $\mathrm{^8Be}$. The lowest-lying $2^-$ resonance state at $E_r = 2.67\,\text{keV}$ provides the dominant baseline to the cross section, while the $3^+$ resonant components at $E_r = 0.33\,\text{MeV}$ and $E_r = 2.66\,\text{MeV}$ influence the spectral shape~\cite{adahchour2003r}. We calculate the sum of each resonance contribution employing the Breit-Wigner formula with a channel radius of $R=5\,\mathrm{fm}$ as recommended by Descouvemont {\it et al.}~\cite{descouvemont2004compilation}.
In panel (c) of Fig.~\ref{fig_resonance}, we demonstrate the dependence of the S-factor $S(E)=\sigma E^{\frac{1}{2}}$ and reaction rate $N_A \langle \sigma v \rangle$ on the Higgs VEV in resonance case 1. Although the resonance peak of the $2^-$ component shifts from 2.67\,keV, it remains the dominant contributor in the low-energy region. Conversely, the shifts in the resonance peaks of the two $3^+$ components modify the shape of the S factor. As shown in the upper-right inset of panel (c), the penetration factor at low neutron energies behaves as $P(E) \sim E^{l+1/2}$. In contrast to the significantly low-energy contributions from the $3^+$ states reported in~\cite{mori2019roles}, we find that the two $3^+$ components with angular momentum $l=1$ do not contribute to the zero-energy S-factor. The shift of cross sections in the Gamow window under BBN temperatures would therefore be governed mainly by the shift of the $2^-$ state.\par

The ratio between forward and reverse reaction rates is governed by the principle of detailed balance. A change in $Q$-value due to a variation in the Higgs VEV would alter the ratios of the forward and reverse reactions, directly modifying the predicted primordial abundances. \par

\section{Results and discussion}
\label{results}
\begin{figure}[ht]
    \centering
    \includegraphics[width=\columnwidth]{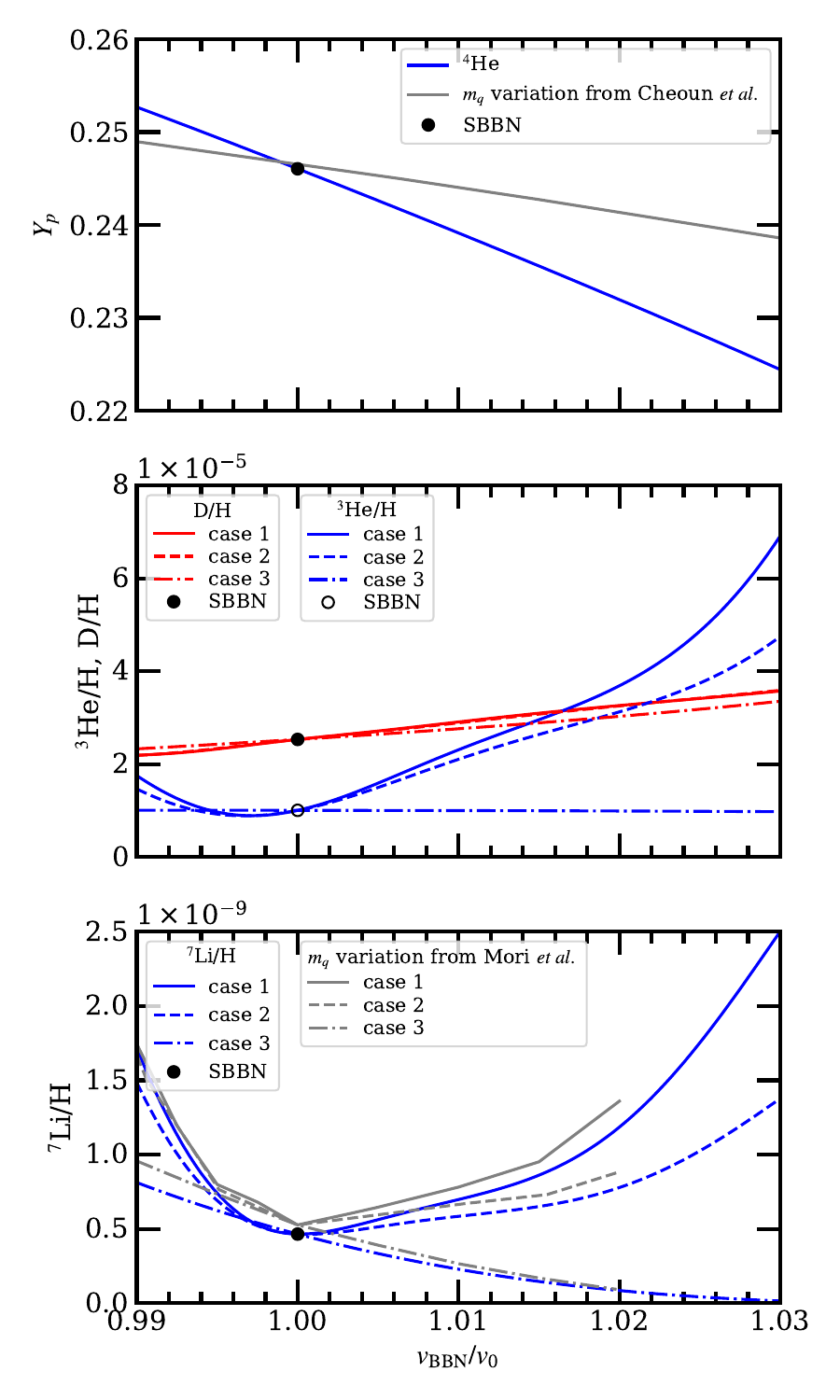}
    \caption{Dependence of primordial abundances on the Higgs VEV. \blue{The $^4$He abundance $Y_p$ is defined as the mass fraction, while D/H, $^3$He/H, and $^7$Li/H represent the number densities relative to hydrogen. 
    } The panels display the abundances of $^4$He (top), D and $^3$He (middle), and $^7$Li (bottom). For comparison, the solid, dashed, and dot-dashed lines correspond to resonance cases 1, 2, and 3, respectively. The gray line in the top panel represents the $^4$He abundance dependence on $m_q$ from Cheoun {\it et al.}~\cite{cheoun2011time}, while the gray line in the bottom panel shows the corresponding $^7$Li dependence from Mori {\it et al.}~\cite{mori2019roles}. \blue{The plotted $^{3}\mathrm{He}$ and $^{7}\mathrm{Li}$ abundances include the contributions from their decay progenitors, $^{3}\mathrm{H}$ and $^{7}\mathrm{Be}$, respectively.}
    }
    \label{fig_BBN_standard_chart_higgs}
\end{figure}

Let us emphasize that we incorporate the effects of the Higgs VEV on both the weak proton-neutron conversion and the thermonuclear reactions into the reaction network. We use a modified version of the Kawano code to calculate BBN~\cite{kawano1992let}. We adopt the neutron lifetime of $\tau=878.4\pm0.5\,\mathrm{s}$ as recommended by the Particle Data Group~\cite{navas2024review}. For the reactions affecting the primordial D abundance, we adopt the most recent reaction rate assessments: $\mathrm{^1H}(n,\gamma)\mathrm{^2H}$ from Chen {\it et al.}~\cite{tbbt-s819}, $\mathrm{^2H}(p,\gamma)\mathrm{^3He}$ from Mossa {\it et al.}~\cite{mossa2020baryon}, and $\mathrm{^2H}(d,n)\mathrm{^3He}$ and $\mathrm{^2H}(d,p)\mathrm{^3H}$ from Pisanti {\it et al.}~\cite{pisanti2021primordial}. All other reaction rates are taken from Descouvemont {\it et al.}~\cite{descouvemont2004compilation}. The dependence of primordial abundances as a function of $v_\mathrm{BBN}$ is shown in Fig.~\ref{fig_BBN_standard_chart_higgs}. Here, we adopt baryon density $\Omega_Bh^2 = 0.022381$ from Planck-ACT-SPT joint analysis~\cite{camphuis2025spt}, corresponding to a baryon-to-photon ratio of $\eta_B=6.12\times10^{-10}$.\par
The primordial $^4$He abundance is more strongly correlated to the Higgs VEV variation compared to the previous result~\cite{cheoun2011time} that introduced only averaged quark mass variation, shown as a gray line in the top panel of Fig.~\ref{fig_BBN_standard_chart_higgs}. This dependence arises from changes in the weak interaction rate: since the $n/p$ ratio is determined by the statistical equilibrium at high temperature, a larger $v_\mathrm{BBN}$ corresponds to a smaller $Q_{np}$ and fewer neutrons. After freeze-out, a larger primordial $v_{\mathrm{BBN}}$ leads to a higher neutron destruction rate. 
The abundance of primordial D increases as increasing $v_{\mathrm{BBN}}$, anticorrelated with the change in the reaction rates of $\mathrm{^1H}(n,\gamma)\mathrm{^2H}$, $\mathrm{^2H}(p,\gamma)\mathrm{^3He}$, $\mathrm{^2H}(d,n)\mathrm{^3He}$, and $\mathrm{^2H}(d,p)\mathrm{^3H}$, as shown in Table~\ref{table_rate_ratio}. 
$^3$He is a crucial ingredient for the synthesis of $^7$Li, as it captures an alpha particle to form $^7$Be, which subsequently decays to $^7$Li via beta emission. We observe that the variation trend of primordial $^7$Li closely follows that of $^3$He. As shown in panel (b) of Fig.~\ref{fig_resonance}, when $v_{\mathrm{BBN}}$ increases, the $\mathrm{^3He}(d,p)\mathrm{^4He}$ resonance peak shifts rightward, suppressing the $^3$He depletion rate at BBN sensitive-temperature \blue{$T_9=1$} so that more $^7\mathrm{Li}$ is produced. Similarly, as $v_{\mathrm{BBN}}$ decreases, the depletion rate drops, leaving more $^3$He that would be converted finally to $^7$Li. This trend is similar to the result of Mori {\it et al.}~\cite{mori2019roles}, as shown by the gray lines in the bottom panel of Fig.~\ref{fig_BBN_standard_chart_higgs}. \par
\begin{figure*}[t]
    \centering
    \subfloat{\includegraphics[width=0.46\textwidth]{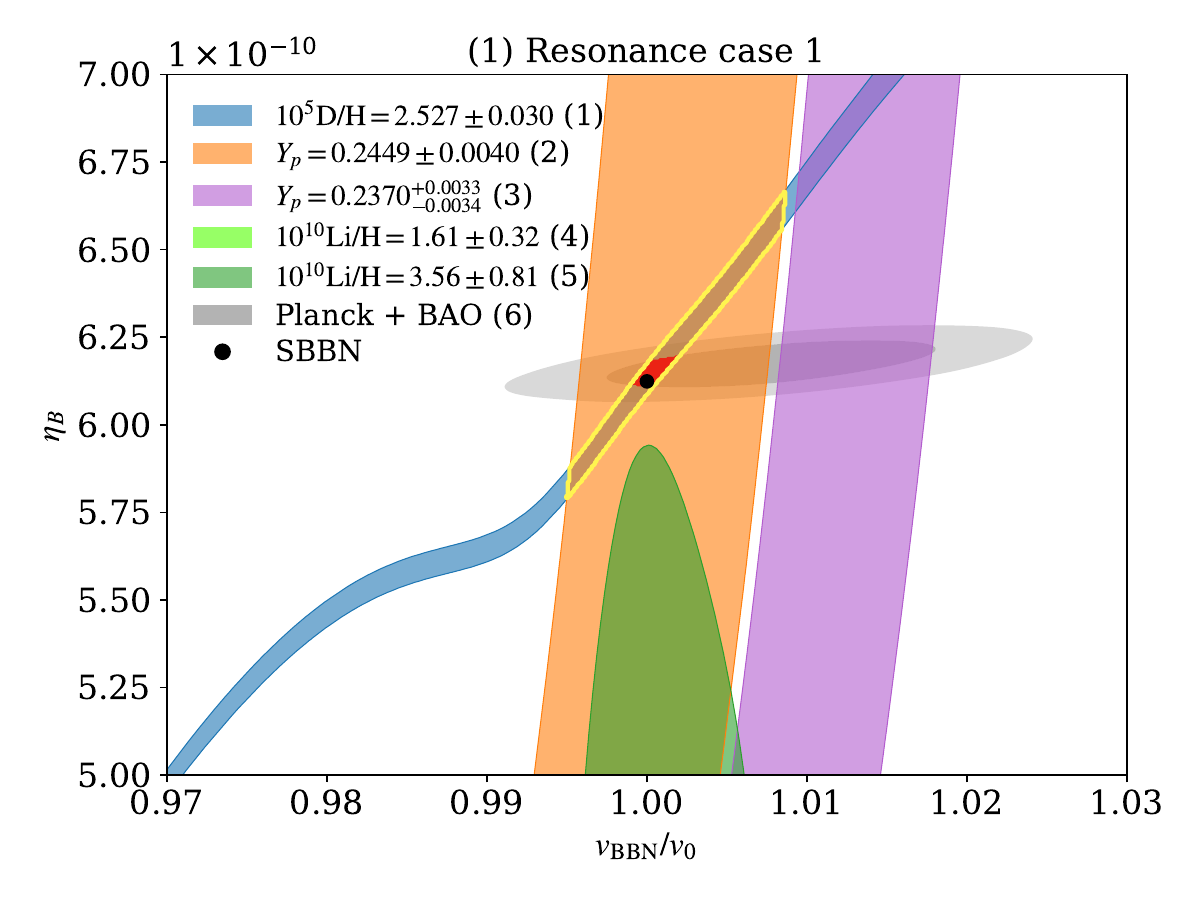}} \\
    \subfloat{\includegraphics[width=0.46\textwidth]{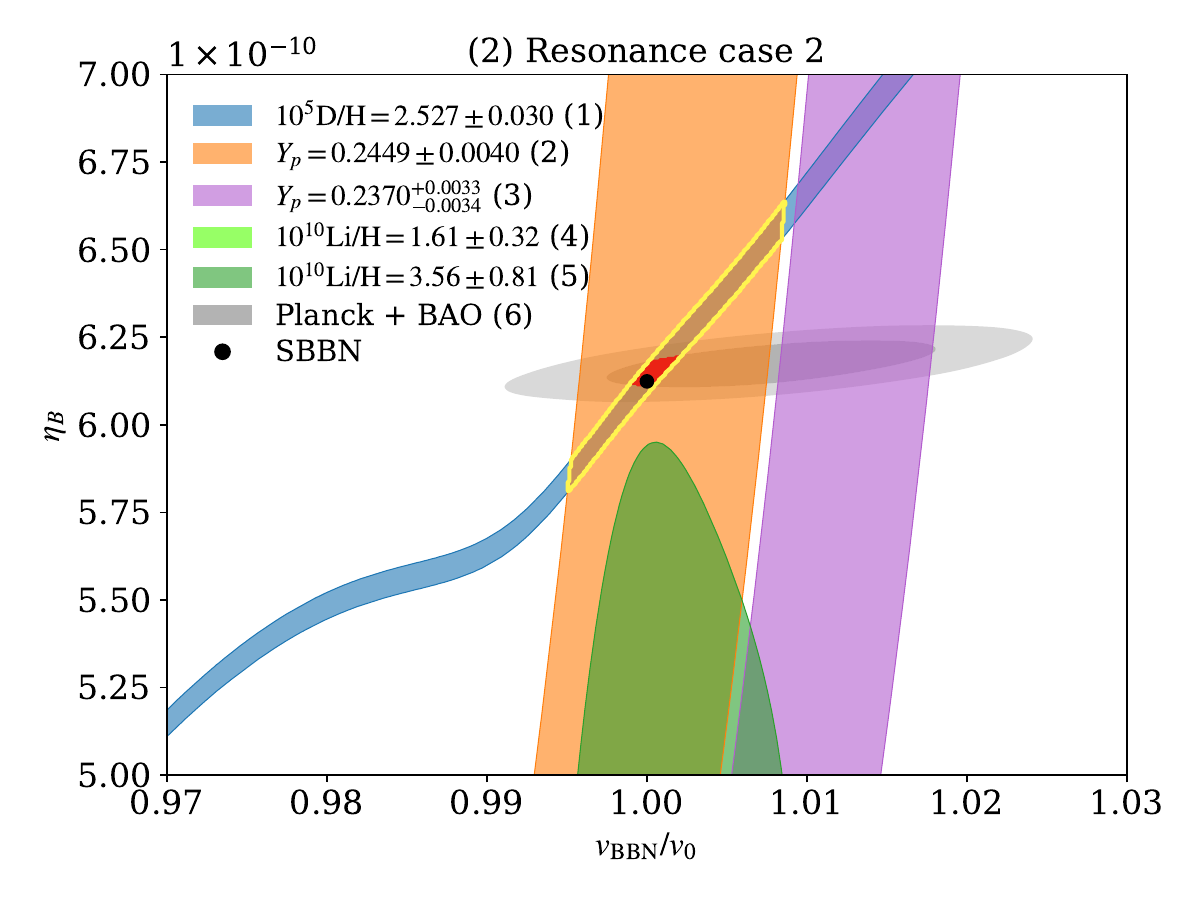}} \\
    \subfloat{\includegraphics[width=0.46\textwidth]{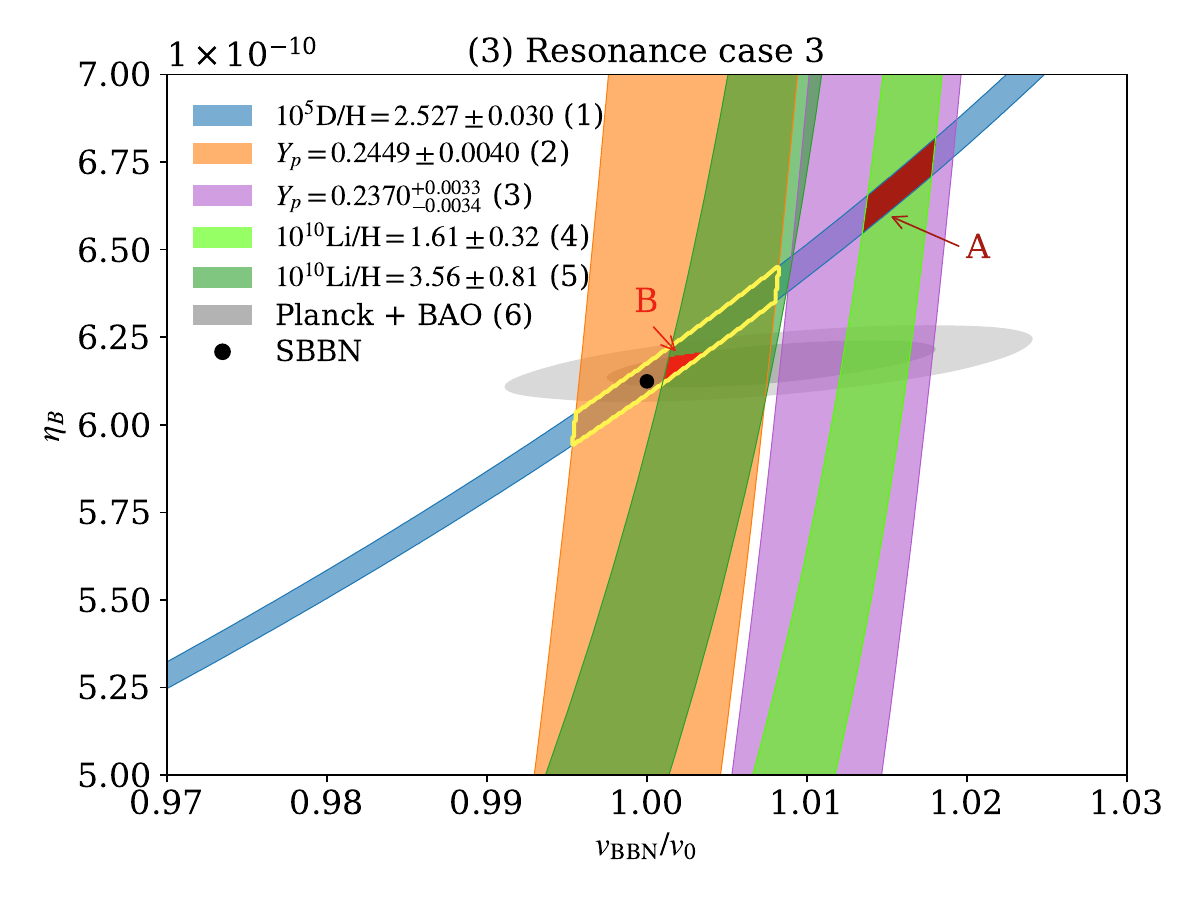}}
    \caption{BBN constraints on the baryon-to-photon ratio $\eta_B$ and Higgs VEV variation. The top, middle, and bottom panels correspond to resonant energy-level cases 1–3. Shaded regions indicate consistency with the following observations: 
    (1) $^4$He from~\cite{aver2015effects} (orange);
    (2) $^4$He from~\cite{matsumoto2022empress} (purple);
    (3) D/H from~\cite{cooke2018one} (blue);
    (4) $^7$Li \blue{Spite plateau} from~\cite{sbordone2010metal} (light green);
    (5) $^7$Li corrected for stellar depletion~\cite{korn2006probable} (dark green);
    (6) Joint constraints from Planck CMB and BAO measurements~\cite{hart2020updated} (gray). \blue{Note that in cases 1 and 2, the light-green Spite-plateau region is absent because a deviation of the Higgs VEV from its present value leads to an increase in the
    primordial $^7\mathrm{Li}$ abundance, as shown in the last panel of Fig.~\ref{fig_BBN_standard_chart_higgs}. Yellow contours across all three panels indicate the $^4\mathrm{He}$~\cite{aver2015effects} and $\mathrm{D}$~\cite{cooke2018one} concordance regions. In case 3, red region A indicates the concordance region among the EMPRESS $^4$He observation~\cite{matsumoto2022empress}, Cooke D/H observation~\cite{cooke2018one} and $^7$Li Spite plateau~\cite{sbordone2010metal}, while red region B indicates the concordance region among the Aver $^4$He observation~\cite{aver2015effects}, Cooke D/H observation~\cite{cooke2018one}, and $^7$Li stellar depletion model~\cite{korn2006probable}.}}
    \label{fig_2d}
\end{figure*}
We further explore the BBN restrictions imposed by observational primordial abundances. Figure~\ref{fig_2d} displays contours constraining the baryon-to-photon ratio $\eta_B$ and Higgs VEV $v_\mathrm{BBN}$ based on observed abundances of D, $^4$He, and $^7$Li. Panels 1 to 3 correspond to resonance cases 1--3, respectively. \par
Primordial D abundance is measured very precisely due to the detailed analysis of multiple Lyman-series absorption lines in quasars by Cooke {\it et al.}~\cite{cooke2018one}, yielding $\mathrm{D/H}=(2.527\pm0.030)\times10^{-5}$. This is shown as a narrow blue band across all three panels. The D abundance exhibits a strong correlation with both baryon density and the Higgs VEV. Since D is virtually never produced in stars, this provides a robust constraint on $\eta_B$. 
Meanwhile, the abundance of $^4$He, as the main product of BBN, also serves as a key constraint on the $\mathrm{\Lambda CDM}$ model. The consistency between BBN calculation and observed $^4$He abundance $Y_p=0.2449\pm0.0040$~\cite{aver2015effects}, shown as orange bands in Fig.~\ref{fig_2d}, has long been regarded as one of the strongest proofs for big bang cosmology. 
We identify a consistent region where a $-0.48\%$ to $+0.86\%$ Higgs VEV deviation and a baryon density of $\eta_B = (6.22\pm0.44)\times10^{-10}$ are allowed, satisfying both $^4$He and D observations, highlighted as yellow contours in Fig.~\ref{fig_2d}.\par

Moreover, $\eta_B$ itself can be independently determined from the CMB power spectrum. However, modifications to the electron mass induced by variations in the Higgs VEV influence the epoch of photon decoupling. Therefore, we utilize the Planck 2018 + BAO constraints on the $\eta_B - m_e$ plane~\cite{hart2020updated}, which account for electron mass variations. The constraint is shown in each panel of Fig.~\ref{fig_2d} as gray contours. Utilizing the $^4$He data from Aver {\it et al.}~\cite{aver2015effects} and D/H observation from Cooke {\it et al.}~\cite{cooke2018one}, we achieve a very tight BBN + CMB joint constraint on $v_\mathrm{BBN}/v_0=1.0010\pm0.0027$ and $\eta_B = (6.16\pm0.05)\times10^{-10}$, as highlighted in red in panels 1 and 2 of Fig.~\ref{fig_2d}.\par

On the other hand, a recent $^4$He survey of extremely metal-poor stars by the EMPRESS group reported $Y_p=0.2370^{+0.0033}_{-0.0034}$, revealing a discrepancy with the SBBN model~\cite{matsumoto2022empress}. This is shown as a violet band in Fig.~\ref{fig_2d}. We need a $\sim 1.5\%$ increase of Higgs VEV and a $\sim 10\%$ increase in $\eta_B$ to satisfy the EMPRESS~\cite{matsumoto2022empress} $^4$He and the Cooke {\it et al.}~\cite{cooke2018one} D constraints. However, this region exhibits a $\sim 4\sigma$ tension with CMB observations.\par

The ``cosmic Li problem'' refers to a $\sim 9\sigma$ difference between the SBBN predictions~\footnote{Since primordial lithium produced in BBN consists of $>99.8\%$ $^7$Li, the calculated $^7$Li abundance is always used to represent the total Li abundance.} and observed $\mathrm{Li}$ abundance in metal-poor halo stars\blue{, i.e., number abundance ratio $\mathrm{Li/H} = (1.6 \pm 0.3) \times 10^{-10}$}~\cite{1982A&A...115..357S, Spite1982Nature, sbordone2010metal}. 
Stellar astrophysics has proposed plausible explanations suggesting that early stars may have destroyed some of the primordial $\mathrm{Li}$~\cite{2015MNRAS.452.3256F,korn2006probable,gruyters2016atomic}, allowing the primordial lithium abundance to reach as high as \blue{$\mathrm{Li/H} = (3.56 \pm 0.81) \times 10^{-10}$}. However, even adopting this abundance constraint, the discrepancy has not been fully resolved. 
In each panel of Fig.~\ref{fig_2d}, these two observational constraints are presented as dark green and light green bands, respectively. As shown in panels 1 and 2, neither of the observed Li abundances agrees with the joint $^4$He + D constraint. This occurs because the $^7$Li abundance tends to increase alongside $^3$He as the Higgs VEV changes, further widening the gap between prediction and observation.\par
In resonance case 3, i.e., when resonance energy level $E_r$ remains unchanged as shown in Fig.~\ref{resonance_illustration}, the reaction rate $^3\mathrm{He}(\alpha,\gamma)^7\mathrm{Be}$ is inversely proportional to the Higgs VEV value, resulting in a lower $^7$Li abundance for a higher $v_\mathrm{BBN}$. Therefore, the concordance regions that satisfy $^4$He, D, and $^7$Li observations could be found. 
Notably, at $\eta_B=(6.68\pm0.12)\times10^{-10}$ and $v/v_0=1.0158 \pm 0.0022$, the newly measured $^4$He abundance by EMPRESS group~\cite{matsumoto2022empress} overlaps with D observations~\cite{cooke2018one} and the Li abundance from metal-poor halo stars~\cite{sbordone2010metal}, highlighted in panel 3 as area A. 
Given the fact that the electron mass scales linearly with $v_\mathrm{BBN}$, this area overlaps with the proposed resolution of the Hubble tension via electron-mass variation, i.e., $m_e/m_{e,0} = 1.0190 \pm 0.0055$~\cite{hart2020updated}. Although area A is in $4\sigma$ tension with the CMB constraint on $\eta_B$, \blue{such a discrepancy might be accommodated in nonstandard cosmological scenarios involving entropy production or an evolution of the baryon-to-photon ratio between the BBN and recombination epochs~\cite{Sobotka:2022vrr,Yeh:2022heq}.}
Furthermore, for the coefficient region $v_\mathrm{BBN}/v_0=1.0022 \pm 0.0012$ and $\eta_B=(6.16\pm0.04)\times10^{-10}$, highlighted as area B in panel 3, the theoretical prediction aligns with the $^4$He observation by Aver {\it et al.}~\cite{aver2015effects}, the D observation by Cooke {\it et al.}~\cite{cooke2018one}, and the Li abundance suggested by Korn {\it et al.}~\cite{korn2006probable} without violating the CMB constraint. This suggests that a $0.22\%$ elevation in the Higgs VEV, in conjunction with stellar depletion in metal-poor stars, could provide a viable resolution to the ``cosmic Li problem.''
Improved stellar depletion models and a better understanding of how nuclear resonances respond to quark mass changes would help strengthen our constraints.

\begin{table*}[t]
\caption{\label{table_constraints} BBN constraints on $\eta_B$ and $v_{\mathrm{BBN}}$.}
\begin{ruledtabular}
\begin{tabular}{lcc}
$^4$He and D constraints (cases 1, 2, and 3) & $\eta_B$ & $v_{\mathrm{BBN}}/v_0$ \\
\hline
Aver $^4$He + Cooke D + Planck + BAO  & $(6.16\pm0.05)\times10^{-10}$ & $1.0010 \pm 0.0027$ \\
Aver $^4$He + Cooke D & $(6.22 \pm 0.44)\times10^{-10}$ & $ 1.0019 \pm 0.0067$ \\
\hline
$^4$He, D and $^7$Li constraints (case 3) & $\eta_B$ & $v_{\mathrm{BBN}}/v_0$ \\
\hline
Matsumoto $^4$He + Cooke D + Sbordone $^7$Li (area A) & $(6.68\pm0.12)\times10^{-10}$ & $1.0158 \pm 0.0022$ \\
Aver $^4$He + Cooke D + Korn $^7$Li + Planck + BAO (area B) & $(6.16\pm0.04)\times10^{-10}$ & $1.0022 \pm 0.0012$ \\

\end{tabular}
\end{ruledtabular}
\end{table*}

Table~\ref{table_constraints} summarizes the constraints on $\eta_B$ and $v_\mathrm{BBN}$, derived from both BBN-independent and combined BBN + CMB data. These findings could serve as a valuable reference for future studies on Higgs VEV variation models in the early universe. \par

\section{Conclusion}
\label{conclusion}
This study investigates the impact of the Higgs VEV on BBN. Variations in the Higgs VEV change the Fermi constant, quark masses, and the electron mass through the electroweak symmetry breaking mechanism and Yukawa couplings. Higgs VEV therefore alters weak-interaction $n\leftrightarrow p$ rates and modifies thermonuclear reaction rates through shifts in reaction $Q$-values and resonance energy levels. 
Through BBN network calculations, we find a strong negative correlation \cyan{between} primordial $^4\mathrm{He}$ \cyan{abundance and} the Higgs VEV, and a weak positive correlation for primordial deuterium abundance. Utilizing observational constraints on these elemental abundance~\cite{cooke2018one,aver2015effects} and Planck 2018 + BAO $\eta_B-m_e$ correlation~\cite{hart2020updated}, we obtain constraints on Higgs VEV and baryon density: $\eta_B = (6.16\pm0.05)\times10^{-10}$ and $v_\mathrm{BBN}/v_0=1.0010 \pm 0.0027$. 
\cyan{Under the assumption that the resonance energy levels do not change with $v_\mathrm{BBN}$, corresponding to case 3 in Fig.~\ref{fig_2d}, we find that a $1.58\%$ increase in the Higgs VEV can bring the predicted primordial D, $^4$He, and $^7$Li abundances into agreement with the deuterium observation~\cite{cooke2018one}, the new EMPRESS $^4$He measurement~\cite{matsumoto2022empress}, and the lithium abundance on the Spite plateau~\cite{sbordone2010metal}, respectively. However, this concordance requires a baryon-to-photon ratio $9.1\%$ higher than the CMB determination, corresponding to an approximately $4\sigma$ tension with the CMB constraint and indicating the need for nonstandard cosmological evolution between the BBN and recombination epochs.}
\cyan{More interestingly, in the CMB-consistent region of parameter space, we find that a $0.2\%$ uplift in the Higgs VEV can reduce the predicted primordial $^7$Li abundance and thereby partially alleviate the cosmic lithium problem. The remaining discrepancy may be accounted for by stellar depletion mechanism.} \par

\vspace{10mm}
\begin{center}
\textbf{ACKNOWLEDGMENTS}
\end{center}
We thank Yu-Cheng Qiu for carefully reading the manuscript and for discussions, and Hongliang Yan, Xiaoting Fu, and Zhi-Yu Zhang for valuable suggestions regarding primordial abundance observations.
This work was partly supported by the National Key R\&D Program of China (No. 2022YFA1602401), the National Natural Science Foundation of China (No. 12335009, No. 12435010, and No. 12325506).
Y. L. is partly supported by the Boya Fellowship of Peking University, the China Postdoctoral Science Foundation under Grant No. 2025T180924 and the Institute for Basic Science of the Republic of Korea under Grants No. IBS-R031-D1.

\end{document}